\documentstyle[12pt,epsf]{article}
\def\fo{\hbox{{1}\kern-.25em\hbox{l}}}
\def\fnote#1#2{\begingroup\def\thefootnote{#1}\footnote{#2}\addtocounter
{footnote}{-1}\endgroup}

\renewcommand{\thefootnote}{\fnsymbol{footnote}}

\def\beq{\begin{equation}}
\def\eeq{\end{equation}}
\def\eq{\end{equation}}
\def\to{\rightarrow}

\def\bsg{\ifmmode B\to X_s\gamma\else $B\to X_s\gamma$\fi}
\def\bsll{\ifmmode B\to X_s\ell^+\ell^-\else $B\to X_s\ell^+\ell^-$\fi}
\def\bstt{\ifmmode B\to X_s\tau^+\tau^-\else $B\to X_s\tau^+\tau^-$\fi}
\def\shat{\ifmmode \hat{s}\else $\hat{s}$\fi}

\newcommand{\newc}{\newcommand}

\newc{\lcal}{\int {\cal L}dt}

\newc{\mstop}{m_{\tilde{t}}}
\newc{\mHpm}{m_{H^\pm}}
\newc{\gsim}{\lower.7ex\hbox{$\;\stackrel{\textstyle>}{\sim}\;$}}
\newc{\lsim}{\lower.7ex\hbox{$\;\stackrel{\textstyle<}{\sim}\;$}}
\newc{\ie}{{\it i.e.}}
\newc{\etal}{{\it et al.}}
\newc{\eg}{{\it e.g.}}
\newc{\kev}{\hbox{\rm\,keV}}
\newc{\mev}{\hbox{\rm\,MeV}}
\newc{\gev}{\hbox{\rm\,GeV}}
\newc{\tev}{\hbox{\rm\,TeV}}
\newc{\xpb}{\hbox{\rm\, pb}}
\newc{\xfb}{\hbox{\rm\, fb}}

\newc{\mtop}{m_t}
\newc{\mbot}{m_b}
\newc{\mz}{m_Z}
\newc{\mw}{M_W}
\newc{\alphasmz}{\alpha_s(m_Z^2)}
\newc{\swsq}{\sin^2\theta_W}
\newc{\tw}{\tan\theta_W}
\newc{\cw}{\cos\theta_W}
\newc{\sw}{\sin\theta_W}
\newc{\BR}{\hbox{\rm BR}}
\newc{\zbb}{Z\to b\bar}
\newc{\Gb}{\Gamma (Z\to b\bar b)}
\newc{\Gh}{\Gamma (Z\to \hbox{\rm hadrons})}
\newc{\rbsm}{R_b^\hbox{\rm sm}}
\newc{\rbsusy}{R_b^\hbox{\rm susy}}
\newc{\drb}{\delta R_b}

\newc{\sgn}{\mbox{sgn}}
\newc{\tbeta}{\tan\beta}
\newc{\uL}{{\tilde u_L}}
\newc{\uR}{{\tilde u_R}}
\newc{\cL}{{\tilde c_L}}
\newc{\cR}{{\tilde c_R}}
\newc{\tL}{{\tilde t_L}}
\newc{\tR}{{\tilde t_R}}
\newc{\dL}{{\tilde d_L}}
\newc{\dR}{{\tilde d_R}}
\newc{\sL}{{\tilde s_L}}
\newc{\sR}{{\tilde s_R}}
\newc{\bL}{{\tilde b_L}}
\newc{\bR}{{\tilde b_R}}
\newc{\eL}{{\tilde e_L}}
\newc{\eR}{{\tilde e_R}}
\newc{\mhp}{m_{H^\pm}}
\newc{\mhalf}{m_{1/2}}

\newc{\lR}{\tilde{l}_R}
\newc{\lL}{\tilde{l}_L}
\newc{\nL}{\tilde{\nu}_L}
\newc{\na}{\chi^0_1}
\newc{\nb}{\chi^0_2}
\newc{\nc}{\chi^0_3}
\newc{\nd}{\chi^0_4}
\newc{\ca}{\chi^{\pm}_1}
\newc{\cb}{\chi^{\pm}_2}
\newc{\camp}{\chi^\mp_1}
\newc{\cbmp}{\chi^\mp_1}
\newc{\capos}{\chi^{+}_1}
\newc{\caneg}{\chi^{-}_1}
\newc{\phit}{\phi_t}
\newc{\phib}{\phi_b}
\newc{\phiew}{\phi_{ew}}
\newc{\htz}{h^0_t}
\newc{\hbz}{h^0_b}
\newc{\hewz}{h^0_{ew}}
\newc{\hsmz}{h^0_{sm}}
\newc{\huz}{h^0_u}
\newc{\hsusyz}{h^0_{susy}}

\def\NPB#1#2#3{Nucl. Phys. B {\bf #1}, #3 (19#2)}
\def\PLB#1#2#3{Phys. Lett. B {\bf #1}, #3 (19#2)}

\def\PRD#1#2#3{Phys. Rev. D {\bf #1}, #3 (19#2)}
\def\PRL#1#2#3{Phys. Rev. Lett. {\bf#1}, #3 (19#2)}

\def\beq{\begin{equation}}
\def\eeq{\end{equation}}
\def\bea{\begin{eqnarray}}
\def\eea{\end{eqnarray}}
\def\slashchar#1{\setbox0=\hbox{$#1$}           
   \dimen0=\wd0                                 
   \setbox1=\hbox{/} \dimen1=\wd1               
   \ifdim\dimen0>\dimen1                        
      \rlap{\hbox to \dimen0{\hfil/\hfil}}      
      #1                                        
   \else                                        
      \rlap{\hbox to \dimen1{\hfil$#1$\hfil}}   
      /                                         
   \fi}                                         %
\catcode`@=11
\long\def\@caption#1[#2]#3{\par\addcontentsline{\csname
  ext@#1\endcsname}{#1}{\protect\numberline{\csname
  the#1\endcsname}{\ignorespaces #2}}\begingroup
    \small
    \@parboxrestore
    \@makecaption{\csname fnum@#1\endcsname}{\ignorespaces #3}\par
  \endgroup}
\catcode`@=12

\def\jfig#1#2#3{
 \begin{figure}
 \centering
 \epsfysize=2.5in
 \hspace*{0in}
 \epsffile{#2}
 \caption{#3}
 \label{#1}
 \end{figure}}

\begin{document}

\begin{titlepage}

\begin{flushleft}
\end{flushleft}
\begin{flushright}
hep-ph/9808287\\
SLAC-PUB-7900 \\
VPI-IPPAP-98-5 \\
July 1998
\end{flushright}
\bigskip



\huge

\begin{center}
{Higgs boson interactions in supersymmetric
theories with large $\tan\beta$}
\end{center}

\large

\vspace{.15in}
\begin{center}

Will Loinaz${}^{a}$ and James D.~Wells${}^b$\fnote{\dagger}{Work
supported by the Department of Energy
under contract DE-AC03-76SF00515.} \\

\vspace{.1in}
{\it ${}^{(a)}$Institute for Particle Physics and Astrophysics \\
Physics Department, Virginia Tech, Blacksburg, VA 24061--0435 \\}
\bigskip
{\it ${}^{(b)}$Stanford Linear Accelerator Center \\
Stanford University, Stanford, California 94309 \\}

\end{center}


\vspace{0.15in}

\begin{abstract}

We show that radiative corrections
to the Higgs potential in supersymmetric theories with large $\tan\beta$
generically lead to large
differences in the light Higgs boson decay branching fractions
compared to those of the standard model Higgs boson.
In contrast, the light Higgs boson production rates are largely unaffected.
We identify $Wh$ associated production followed by Higgs boson
decays to photons or to leptons via $WW^*$ as potential experimental probes
of these theories.
\bigskip

\begin{center}
(Submitted to Phys. Lett. B)
\end{center}

\end{abstract}

\end{titlepage}

\baselineskip=18pt




\vfill
\eject


It is well known that supersymmetry requires two Higgs doublets to give
masses to the up-type quarks and to the down-type quarks.  Hence, we
use the terminology ``up-Higgs'' and ``down-Higgs'' to indicate these
two Higgs bosons, $H_u$ and $H_d$ respectively.
The ratio of the vacuum expectation values,
\beq
\tan\beta  \equiv  \frac{\langle H^0_u\rangle}{\langle H^0_d\rangle},
\eeq
has both theoretical and experimental consequences.

Theoretically,
a large value of $\tan\beta$ near $m_t/m_b$ appears to be required in
simple $b-\tau -t$ Yukawa unification theories such as supersymmetric
$SO(10)$~\cite{hall94:7048}.
Since the b-Yukawa and the t-Yukawa couplings are equal at
the high scale they will be nearly equal at the low scales.  This is because
the bottom and top quarks are distinguished only by their weak hypercharge,
and renormalization group evolution is dominated by large Yukawa and QCD
interactions.

Furthermore, a large value of $\tan\beta\gsim 30$ is required for minimal
gauge mediated models which solve the soft-CP
problem~\cite{dine95:1362,babu96:3070}.
That is,
CP violation effects are non-existent
in the soft mass terms of a softly broken supersymmetric
theory with gauge mediated supersymmetry breaking and with $B_\mu(M)=0$
at the messenger scale $M$.  If $B_\mu$ is zero at the messenger scale
then arbitrary phases in the lagrangian can be rotated away, and CP violating
effects do not get induced by renormalization group running.
However, renormalization group running does induce a non-zero (CP conserving)
$B_\mu$ term at the weak scale which is phenomenologically viable only
if $\tan\beta$ is large~\cite{babu96:3070,rattazzi97:297}.
This can be seen readily from one of the
conditions for proper electroweak symmetry breaking:
\beq
\sin\beta\cos\beta = \frac{B_\mu(M)+\Delta B_\mu}{m^2_A}.
\eeq
Since $B_\mu (M)=0$ and $\Delta B_\mu$ is a small renormalization group
induced mass, and $m_A$ must be large enough to escape detection, the
right-hand side of the equation is much less than 1. Therefore,
\beq
\sin\beta\cos\beta \ll 1 ~~~\Longrightarrow ~~~ \tan\beta \gg 1.
\eeq

Unification of the third family Yukawa couplings and CP conservation
in gauge mediated models both imply a strong preference for large
$\tan\beta$. It is important to keep in mind that these preferences
survive even when there is non-universality among the scalar masses
in the theory.  In supergravity,
non-universality is even expected. Furthermore, large $\tan\beta$
is of course accessible in parameter space even if third family
Yukawa unification is not realized, or if the CP violating interactions
are kept under control in a supersymmetric theory by some other means.


The mass matrix of the scalar Higgs bosons in the $\{H^0_d,H^0_u\}$
basis can be parametrized by,
\beq
M^2= \left( \begin{array}{cc}
  m_A^2\sin^2\beta+m^2_Z\cos^2\beta & -\sin\beta\cos\beta (m_A^2+m_Z^2) \\
 -\sin\beta\cos\beta (m^2_A+m_Z^2) & m_A^2\cos^2\beta +m^2_Z\sin^2\beta
 \end{array} \right) +
 \left( \begin{array}{cc}
  A_{11} & A_{12} \\
  A_{12} & A_{22}
 \end{array} \right) ,
\eeq
where $m_A$ is the physical pseudoscalar Higgs boson mass, and $A_{ij}$
are radiative corrections induced by particle and sparticle loops.
The dependences of $A_{ij}$ on the sparticle masses and mixing angles
can be found in several references~\cite{okada91:1}.
For larger $\tan\beta$ and arbitrary mixing, a random scan over
supersymmetric parameter space is enough
to convince oneself
that the off diagonal $A_{12}$ can be as large as $\pm (40\gev)^2$.

Models with large $\tan{\beta}$ and large
$m_A$ allow suppression of the $b\bar b$ decay mode resulting from 
cancellation of tree-level and off-diagonal terms in the Higgs
mass matrix.
There are two expansions that we would like
to perform to illustrate this feature.  The first is
an expansion of the Higgs mass matrix about pure $H_u^0$ and $H^0_d$ {\it mass}
eigenvalues.  For
these weak eigenstates to be mass eigenstates, mixing
in the Higgs mass matrix is not allowed.  At tree level this is impossible since
in our convention $M^2_{12}$ is negative definite.  However, when radiative
corrections are introduced a cancellation can occur between the
tree-level off-diagonal piece and the quantum corrections.  For this
to occur,
\beq
 -\sin\beta\cos\beta (m^2_Z+m^2_A)+A_{12} =0
\eeq
is required.  Note that if $\tan\beta$ is very large then the
tree-level contribution gets small and the radiative correction
term, $A_{12}$, can compete with it to yield $M_{12}=0$~\cite{baer98:4446}.
Solving for
the mass eigenvalues we obtain,
\bea
m^2_A & = & -m^2_Z+\frac{A_{12}}{\sin\beta\cos\beta} \\
m^2_{H^0_d} & = & m^2_Z\cos 2\beta +A_{12}\tan\beta + A_{11} \\
m^2_{H^0_u} & = & -m^2_Z\cos 2\beta +\frac{A_{12}}{\tan\beta}+A_{22}.
\eea

To demonstrate the behavior of these eigenvalues, we plot their values 
in the limit that all soft squark masses are
the same ($m_{susy}$)
at the low scale.  Squark mass degeneracy is broken by mixing introduced
through a common $A$-term
at the low scale and the $\mu$ term.  Taking $\tan\beta=40$ and
$\mu = A=300\gev$
we plot $m_A\simeq m_{H^0_d}$, $m_{H^0_u}$ and
$m_A+m_{H^0_d}$ as a function of $m_{susy}$.  We have plotted
$m_A+m_{H^0_d}$ since it is constrained to be greater than
$\sim 142\gev$ from LEPII analyses~\cite{LEPII}.
\jfig{eigs}{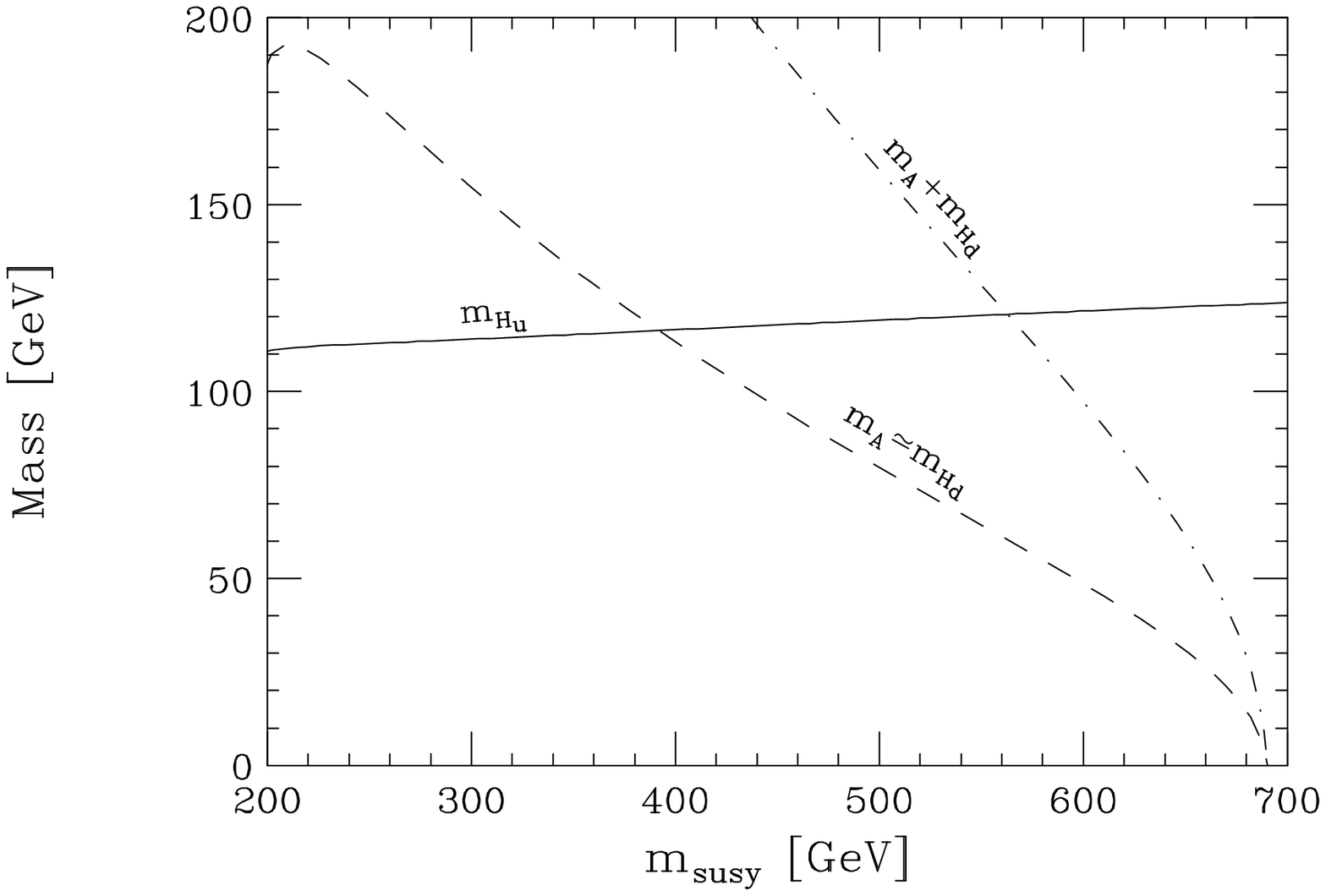}{The mass eigenvalues $m_A\simeq m_{H^0_d}$,
$m_{H^0_u}$, and $m_A+m_{H^0_d}$
as a function of $m_{susy}$.  $m_A+m_{H^0_d}$
is constrained to be greater than $\sim 142\gev$ from LEPII analyses.}
We see that $H^0_d$ and $A$ have mass,
\beq
m_A^2 \simeq m^2_{H^0_d} \simeq -m_Z^2 + A_{12} \tan\beta
\eeq
which ranges widely with $m_{susy}$ in this model.  For $A_{12}$
not too small $m_A$ is naturally large for large $\tan\beta$.
$H^0_u$ has a mass equal to,
\beq
\label{Hu mass}
m^2_h = m^2_{H^0_u} \simeq m^2_Z+A_{22},
\eeq
which is logarithmically sensitive to $m_{susy}$.

The $H^0_u$ mass eigenvalue is always light (between $m_Z$ and $135\gev$).
By assumption it cannot decay into $b\bar b$ or $\tau^+\tau^-$, and is
only allowed to decay
into $c\bar c$, $gg$, $\gamma\gamma$, $WW^*$, $Z^{(*)}\gamma$,
and $ZZ^{(*)}$.  Since $H^0_u$ carries all the burden of electroweak
symmetry breaking, it couples with full
electroweak strength  to the vector bosons, and the up-type quarks.
Therefore, $H^0_u$ acts like the standard model Higgs in this limit in
every way
except it does not couple to down quarks or leptons\cite{dmb comment}.
The decay branching fractions~\cite{djouadi98} for this state are shown in
Fig.~\ref{Hu decays}.
\jfig{Hu decays}{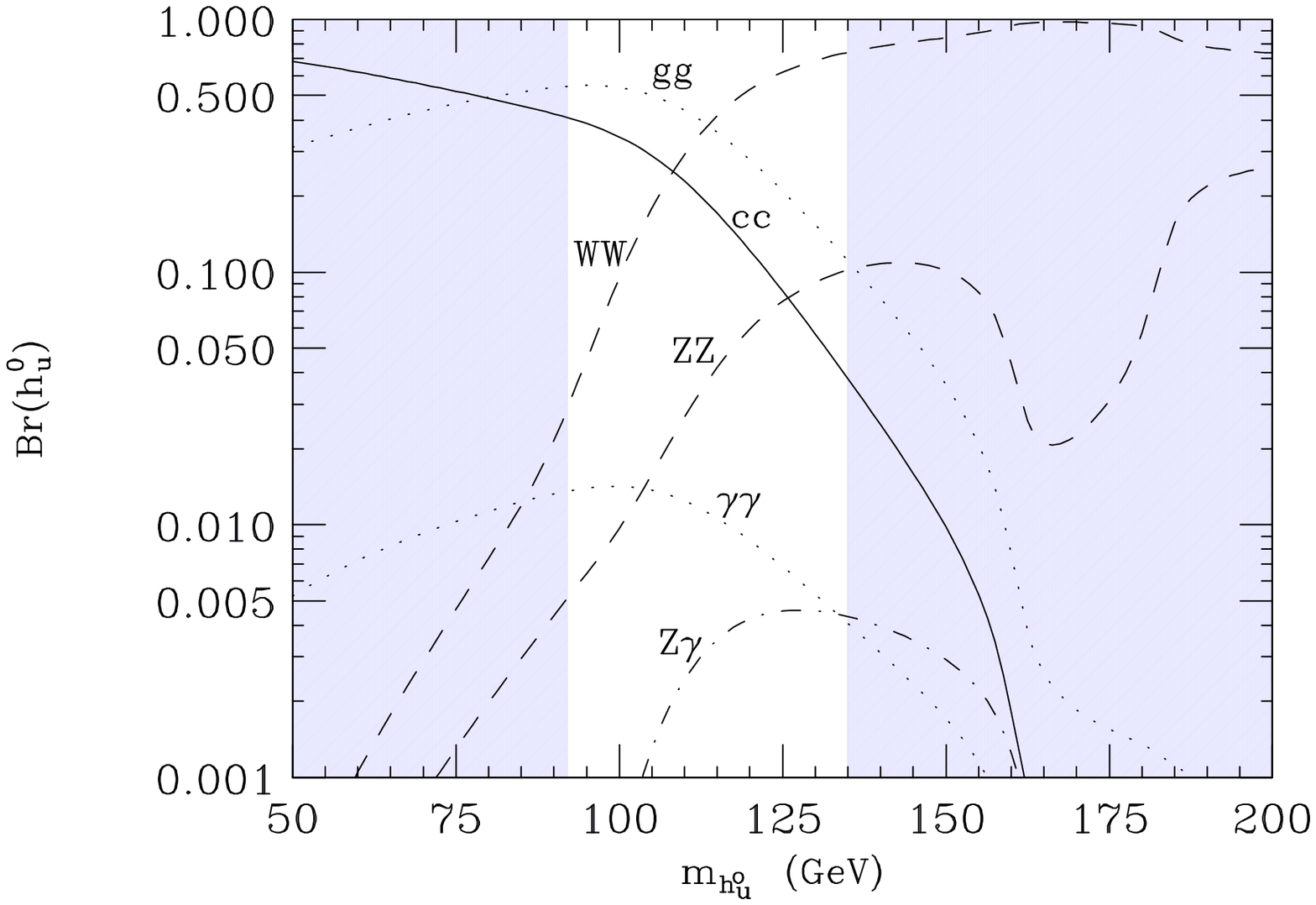}{Decay branching fractions of the $H_u^0$ mass
eigenstate. The shaded region is theoretically not allowed
with high $\tan\beta$.}
According to Eq.~\ref{Hu mass}, the shaded region is theoretically not allowed
in the high $\tan\beta$ limit.  Only in the region 
$m_Z\lsim m_{H^0_u}\lsim 135\gev$ does one expect to find the $H^0_u$
at a high energy collider.

The above discussion may appear unsatisfactory since it seems to imply
that a special finetuning of $m_A$ is needed for interesting effects
to occur.  To address this concern, we expand in a second limit, the
large $m_A$ limit, in order to justify our claim that large deviations
occur generically in large $\tan\beta$ and large $m_A$.

For a moment, let us ignore the radiative corrections and expand about
large $m_A$ with $A_{ij}=0$.  The lightest Higgs boson coupling
to quarks and vector bosons compared to the standard model Higgs boson coupling
to quarks and vector bosons is,
\bea
\frac{(h\bar d d)_{susy}}{(h\bar d d)_{sm}}
  & = & 1+2\epsilon_A + 2\epsilon^2_A +\cdots \\
\frac{(h\bar u u)_{susy}}{(h\bar u u)_{sm}}
  & = & 1-2\epsilon_A\epsilon_\beta^2-4\epsilon_A^2\epsilon_\beta^2+\cdots \\
\frac{(hVV)_{susy}}{(hVV)_{sm}}
  & = & 1-2\epsilon_A^2 \epsilon_\beta^2 +\cdots
\eea
where
\beq
\epsilon_A\equiv m^2_Z/m^2_A ~~~{\rm and}~~~\epsilon_\beta \equiv 1/\tan\beta .
\eeq

There are several points to notice in the above expressions.  First,
the coupling to up-type fermions decouples to the standard model value much
faster than the coupling to the down-type fermions by a factor of
$1/\tan^2\beta$.  Second, the vector boson couplings to the light Higgs
boson decouples even more rapidly to the standard model result.  Therefore,
even at tree-level, we expect the coupling to down-type fermions to deviate
from the standard model much more than the couplings to the up-type fermions
and vector bosons.

Now, if we add radiative corrections to the expansion we get an even
more interesting result.  Keeping the first tree-level correction and
the leading radiative correction terms,
\bea
\frac{(h\bar d d)_{susy}}{(h\bar d d)_{sm}}
  & = & 1+2\epsilon_A -\frac{A_{12}}{m^2_Z}\frac{\epsilon_A}{\epsilon_\beta}
        +\cdots \\
\frac{(h\bar u u)_{susy}}{(h\bar u u)_{sm}}
  & = & 1-2\epsilon_A \epsilon_\beta^2 -\frac{1}{2}\frac{A^2_{12}}{m^4_Z}
        \epsilon_A^2+\frac{A_{12}}{m^2_Z}\epsilon_A\epsilon_\beta + \cdots \\
\frac{(hVV)_{susy}}{(hVV)_{sm}}
  & = & 1-2\epsilon_A^2\epsilon_\beta^2 -\frac{1}{2}\frac{A^2_{12}}{m^4_Z}
       \epsilon_A^2 +\cdots .
\eea
In this case the up-type fermions and vector boson couplings to the
light Higgs boson still decouples rapidly to the standard model result.
In large $\tan\beta$ models, $A_{12} \sim m^2_Z$ for much of the 
parameter space.  Further, unless $m_A$ is very heavy, we expect 
$\epsilon_A$ to be less than or comparable to $\epsilon_{\beta}$
for large $\tan\beta$.  In such models there is an ${\cal O}(1)$ correction
to the down-type fermion couplings to the light Higgs boson.  In particular,
we see that if $A_{12}\simeq m^2_Z\epsilon_A/\epsilon_\beta$ then 
$h=H^0_u$ and the coupling
to down-type fermions may be shut off completely.  This is the same result
we concluded from the previous expansion above.  Here, however,
we have shown that
the coupling to down-type fermions is generically altered by ${\cal O}(1)$
corrections in either direction depending on the sign of $A_{12}$,
but the coupling to up-type fermions and vector bosons are
standard model-like.


At high $\tan\beta$, the pseudoscalar state $A^0$ may be produced 
copiously at high energy colliders, and significant constraints
obtained~\cite{mA at high tanbeta}.  Here we focus on the
CP-even scalar states, since one, which we are calling $h$, is guaranteed
to be light (below $\lsim 135\gev$).
The main production mechanisms for $h$ at high energy
hadron colliders are $qq'\to Wh$, $gg\to h$, and $qq\to t\bar t h$.
At electron-positron colliders the production modes are $e^+e^-\to hZ$,
$e^+e^-\to t\bar t h$, and $\gamma\gamma \to h$.
Since the $h$ couplings to the top-quark and vector bosons are very
close to the standard model values, these cross-sections are not
expected to deviate much from the Standard Model.  This is good news
to not have to worry about complicated mixing between
non-standard production and decay.  Only the branching fractions are altered in
high $\tan\beta$ models\footnote{Note that $\mu^+\mu^-\to h$ would
be altered at a muon collider.}.

For a Higgs boson mass near $120\gev$ and below, the main decay modes
are $h\to b\bar b$ and $h\to \tau\tau$.  If these two were the only
decay modes then deviations in the down-type fermion couplings would
just cancel in the branching fraction and only the standard model result
would be seen.  However, there are other important decay modes of the
Higgs boson, most notably the $h\to \gamma\gamma$ decay mode.  In the
standard model its branching fraction is  $\sim 1.4\times 10^{-3}$
in the mass range $100\gev < m_h < 150\gev$.  In supersymmetry,
this branching fraction can be altered by three factors.  First, supersymmetry
sparticle loops involving squarks and charged Higgs bosons can change
the $h\to \gamma\gamma$ amplitude. Given current limits on sparticle
masses and some fortuitously large loop integrals for the $W$ boson
compared to charged scalars, this effect is small over almost all of
the parameter space~\cite{kane96:213}.
Second, changes in the $h\bar t t$ and $hWW$ couplings
can change the $h\to \gamma\gamma$ amplitude.  However, we argued above
that this does not occur in large $\tan\beta$ models.  And third,
large corrections to $h\bar b b$ and $h\tau\tau$ couplings can change
the denominator in $B(h\to \gamma\gamma)$.  This effect is large
in the high $\tan\beta$ models.

Since $\Gamma (h\to WW^*)$ is unaffected in the large $\tan\beta$ region also,
its branching ratio is determined only by the changes wrought by deviations
in the
down-type fermion couplings.
In Fig.~\ref{xi}
\jfig{xi}{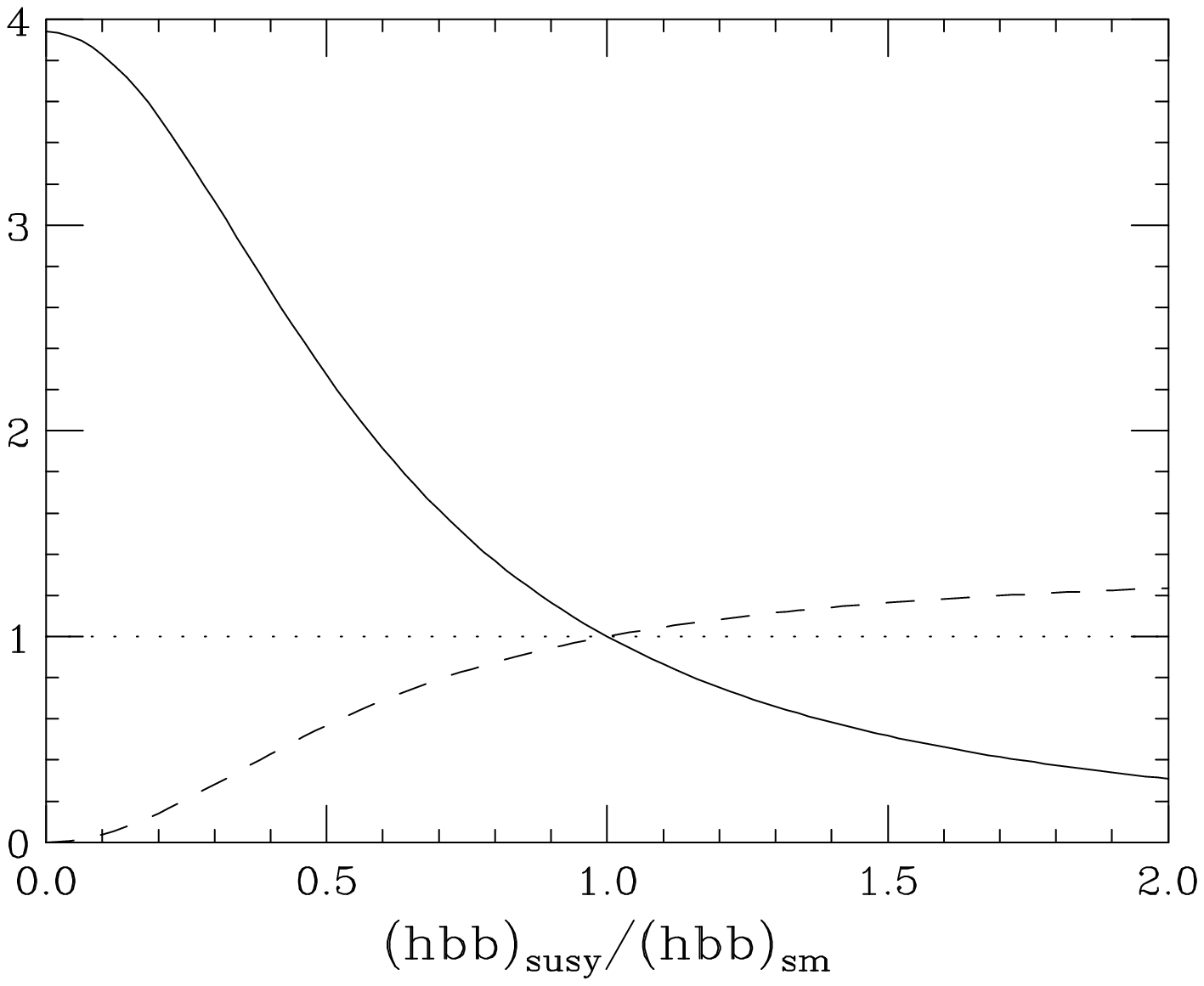}{$B(h\to VV)_{susy}/B(h\to VV)_{sm}$ (solid) and
$B(h\to b\bar b)_{susy}/B(h\to b\bar b)_{sm}$ (dash) as a function
of the variation in the $h\bar b b$ coupling for $m_h=120\gev$.
A dotted line at the
value 1 corresponds to the standard model reference.}
we plot the deviation in the $h\to VV$ ($V=\gamma , W^{(*)}, Z^{(*)}$)
branching fraction compared to the
standard model branching fraction as a function
of $(hbb)_{susy}/(hbb)_{sm}$.  We also plot the deviation of 
$B(h\to b\bar b)$ for reference.  The plot was
made with $m_h=120\gev$.  Since the Higgs boson decays predominantly
into down-type fermions at this mass, the branching fraction into
vector bosons is most sensitive to deviations in the $hbb$ coupling.
The sensitivity (insensitivity)
increases even more for $B(h\to VV)$ ($B(h\to b\bar b)$) for lighter Higgs
masses.

The high $\tan\beta$ gauge mediated models with $B_\mu =0$ at the
messenger scale provide a good specific illustration of the expected
deviations.  It is conservative since the value of $\mu$ and $m_A$
tends to be higher in these high $\tan\beta$ models than supergravity
mediated theories.    If we take the messenger scale to be equal to
$10^4\Lambda$, where $\Lambda\equiv F_S/S$ (see~\cite{babu96:3070}
for discussions
of gauge mediation parameters), then we can self-consistently calculate
the value of $\tan\beta$ as a function of $\Lambda$ which ensures
that $B_\mu (M)=0$.  We find that with
chargino mass between $100\gev$ and $200\gev$,
\bea
& 105\gev \lsim m_h \lsim 115\gev & \\
& 1.06 \lsim \frac{B(h\to b\bar b)_{susy}}{B(h\to b\bar b)_{sm}} \lsim 1.03 \\
& 0.55 \lsim \frac{B(h\to VV)_{susy}}{B(h\to VV)_{sm}} \lsim 0.85 .
\eea
As expected the deviation in the Higgs to vector bosons branching
fraction is significant.  We have also scanned parameter space in
supergravity theories with non-universal scalar masses and found
$\pm {\cal O}(1)$ deviations in the $h\to VV$ branching fractions over
much of parameter space.

An important question is whether or not experiment can see the effects
of these deviations in the Higgs branching fractions to vector bosons.
It is probably hopeless to try to see even $50\%$ effects in
the $\gamma\gamma$ rate from $gg\to h$ fusion. The QCD corrections to
this rate are nearly a factor of two by themselves with
large uncertainties~\cite{spira98:203}.
A more fruitful approach may be to search
for $Wh$ associated production with the $W$ decaying leptonically and
the Higgs boson decaying to
$\gamma\gamma$~\cite{gunion91:510} or
$WW^*\to l\nu l\nu$~\cite{wells97:1504,baer98:4446}.
The total rates for these production mechanisms in the standard model
at the LHC with $\sqrt{s}=14\tev$
is given in Fig.~\ref{vvprod}.
\jfig{vvprod}{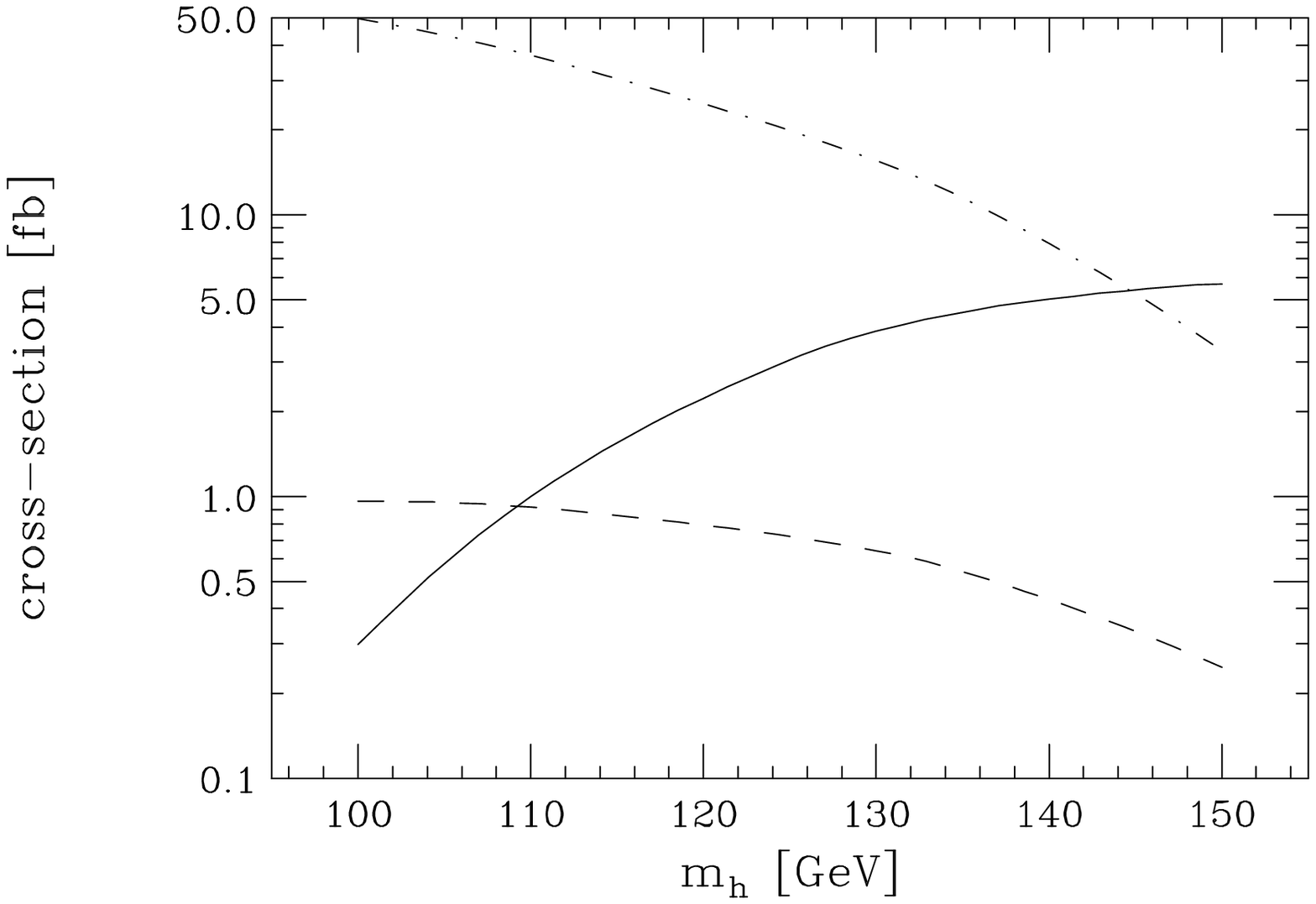}{$\sigma (Wh\to 3l)$ (solid),
$\sigma(Wh\to l\gamma\gamma)$ (dash), and $\sigma (Wh\to l\tau\tau)$
(dot-dash) in the standard model at the LHC with $14\tev$ center-of-mass
energy.}
The advantages to these modes is the smaller background and the ability
to cancel some systematics by taking ratios of this with some other
electroweak process, for example, $\sigma (WZ)$.

The maximum possible sensitivity to a deviation in the cross-section can
be estimated by assuming a perfect detector with no background and
no theoretical uncertainty in the calculation of the standard model rate.
The sensitivity to deviations in the branching fraction are
then given by $\Delta B/B\geq 1/\sqrt{\sigma I}$ 
where $I$ is the time integrated
luminosity, and $\sigma$ is the specific cross-section we are
investigating (e.g., $\sigma(l\gamma\gamma)$).
For $100\xfb^{-1}$ this sensitivity is {\it at best}
$10\%$ at the $1\sigma$ level.
At $1000\xfb^{-1}$, which would require several years of high-luminosity LHC
running, the sensitivity is at best $3\%$ at the $1\sigma$ level.
The ATLAS and CMS technical design reports~\cite{tdr} provide estimates
on signal efficiencies and background rates~\cite{tdr}
(both real and fake) to this
process for the ATLAS and CMS detectors that will be installed at LHC.
An updated study at Snowmass~96 estimated that with LHC data and
NLC data ($\sqrt{s}=500\gev$ and $200\xfb^{-1}$) one could determine
$B(h\to \gamma\gamma)$ to $\lsim 16\%$ accuracy~\cite{snowmass} in the
range $80\lsim m_h\lsim 130\gev$.

Such a measurement of $B(h\to\gamma\gamma)$ would be
advantageous in piecing together large $\tan\beta$ supersymmetric theories.
This is even true in the gauge mediated model discussed above.
Gauge mediation models do not allow much mixing in the top squark sector
because the $A$-terms are zero at the messenger scale.  Therefore, the
Higgs masses in this scenario were rather light, and the off-diagonal
$A_{12}$ entry in the Higgs mass matrix was not as large as it generically
is in supergravity models.  In the case of large mixing, the $hbb$ coupling
can be more volatile than we found in gauge mediation models and
the Higgs mass can be larger.  In Fig.~\ref{vvprod} we see that the
$l\gamma\gamma$ rate falls off rather rapidly when $m_h$ is above $120\gev$,
and the $Wh\to 3l$ rate increases rapidly.  For Higgs masses above
$120\gev$ the most useful probe of Higgs coupling deviations may very well
be the $3l$ signal~\cite{wells97:1504,baer98:4446}.
Here, the background could be as large
as $1\xfb$~\cite{baer98:4446}.
With the detailed background cuts applied also to the signal we estimate that
for a $130\gev$ Higgs boson the deviation in the $3l$ signal
at best could be measured to within $20\%$ at $1\sigma$ with $100\xfb^{-1}$.
With $10^4\xfb^{-1}$ the rate could be measured to perhaps better than $8\%$.
Therefore, it is extremely important, and even necessary to study the
$3l$ signal in detail at the LHC as a probe of non-standard couplings
to the light Higgs boson.  Interestingly, the $3l$ signal from supersymmetric
chargino and neutralino production is highly suppressed in the large
$\tan\beta$ region~\cite{hightanbeta}.
The $Wh\to 3l$ signal therefore dominates the $3l$ signal
from superpartner production.

It should be noted in conclusion that
the gauge mediation model which we used to calculate these results is
correlated with deviations in the $b\to s\gamma$
observable~\cite{rattazzi97:297}. Substantial deviations in $b\to s\gamma$
are generically expected in all large $\tan\beta$ models. However, given
the expected supersymmetry contribution to this amplitude, the expected
experimental data, and the uncertainties in QCD corrections, it is likely
that deviations will show up as anomalies at the few $\sigma$
level at $B$-factories, and so conclusive evidence for new physics would
not be ensured.
A suppression of $l\gamma\gamma$ events at LHC for the
given mass range would then be helpful to support the claim of supersymmetric
theories with high $\tan\beta$. Furthermore, there are large regions of
parameter space in the MSSM not necessarily tied to any minimal model
which predicts small deviations in $B(b\to s\gamma)$ but
${\cal O}(1)$ corrections to the $h\bar bb$ coupling.   These are best
investigated by the Higgs boson observables discussed above.




\begin{thebibliography}{20}
\bibitem{dmb comment}
Finite $\Delta m_b$ corrections allow $H_u^0$ to decay to $\bar bb$.
However, small mixing between $H^0_u$ and $H^0_d$ can yield an eigenvalue
$h$ which does not couple to $\bar bb$, and has an insignificant coupling
to $\tau^+\tau^-$.
\end{thebibliography}
\end{document}